\def\year{2022}\relax
\documentclass[letterpaper]{article} 
\usepackage{aaai22}  
\usepackage{times}  
\usepackage{helvet} 
\usepackage{courier}  
\usepackage[hyphens]{url}  
\usepackage{graphicx} 
\usepackage{natbib}  
\usepackage{caption} 
\DeclareCaptionStyle{ruled}{labelfont=normalfont,labelsep=colon,strut=off} 
\usepackage{amsmath}
\usepackage{dcolumn}
\usepackage{kotex}
\usepackage{color}
\usepackage{hhline}
\usepackage{verbatim}
\usepackage{multirow}
\usepackage{amssymb}
\usepackage{rotating}
\usepackage{subcaption}
\usepackage{dblfloatfix}
\usepackage{enumitem}
\usepackage{float}
\usepackage[dvipsnames]{xcolor}
\usepackage{booktabs}

\usepackage[switch]{lineno} 
\begin{document}

\title{Learning Economic Indicators by Aggregating \\ Multi-Level  Geospatial Information
}

\author{
    Sungwon Park\thanks{Equal contribution to this work.}\textsuperscript{\rm 1},
    Sungwon Han\footnotemark[1]\textsuperscript{\rm 1},
    Donghyun Ahn\textsuperscript{\rm 1},
    Jaeyeon Kim\textsuperscript{\rm 2},
    Jeasurk Yang\textsuperscript{\rm 3}, \\
    Susang Lee\textsuperscript{\rm 1},
    Seunghoon Hong\textsuperscript{\rm 1},
    Jihee Kim\textsuperscript{\rm 1},
    Sangyoon Park\textsuperscript{\rm 4},
    Hyunjoo Yang\textsuperscript{\rm 5},
    Meeyoung Cha\textsuperscript{\rm 6,1} \\
}

\affiliations{
    \textsuperscript{\rm 1}KAIST \mbox{} 
    \textsuperscript{\rm 2}New York University \mbox{} 
    \textsuperscript{\rm 3}National University of Singapore \mbox{}   \\ 
    \textsuperscript{\rm 4}University of Hong Kong \mbox{} 
    \textsuperscript{\rm 5}Sogang University
    \textsuperscript{\rm 6}Institute for Basic Science \mbox{} \\
    
    \{psw0416, lion4151, segaukwa, susang88, seunghoon.hong, jiheekim\}@kaist.ac.kr, \\ jk7362@nyu.edu,~~jeasurk91@u.nus.edu,~~sangyoon@hku.hk,~~hyang@sogang.ac.kr,~~mcha@ibs.re.kr
  
}

\newcommand{\img}{{\bf d}}
\newcommand{\district}{D}
\newcommand{\districtdata}{\mathcal{D}}

\maketitle
\begin{abstract}
High-resolution daytime satellite imagery has become a promising source to study economic activities. These images display detailed terrain over large areas and allow zooming into smaller neighborhoods. Existing methods, however, have utilized images only in a single-level geographical unit. This research presents a deep learning model to predict economic indicators via aggregating traits observed from multiple levels of geographical units. The model first measures hyperlocal economy over small communities via ordinal regression. The next step extracts district-level features by summarizing interconnection among hyperlocal economies. In the final step, the model estimates economic indicators of districts via aggregating the hyperlocal and district information. Our new multi-level learning model substantially outperforms strong baselines in predicting key indicators such as population, purchasing power, and energy consumption. The model is also robust against data shortage; the trained features from one country can generalize to other countries when evaluated with data gathered from Malaysia, the Philippines, Thailand, and Vietnam. We discuss the multi-level model's implications for measuring inequality, which is the essential first step in policy and social science research on inequality and poverty.
\end{abstract}

\section{Introduction}

A recent hike in the availability of high-resolution daytime satellite imagery has revolutionized how we collect data and measure economic indicators. These images display detailed land cover, and they are comparatively easy to collect on a large scale. Such abundance enables satellite imagery to be applied as raw input for sophisticated predictions using deep learning. For example, studies have utilized a convolutional neural network-based framework on satellite images to estimate population~\cite{robinson2017deep}, poverty~\cite{jean2016combining}, and other economic indicators~\cite{han2020lightweight}.

Conventional methods have largely tackled information processing at a single geographic level. For example, some have utilized information at the \textbf{hyperlocal-level} that covers small geographic areas like neighborhoods~\cite{pandey2018multi,han2020learning}, whereas others have used information at the \textbf{district-level} that cover larger administrative units like counties and cities~\cite{jean2016combining}. Models utilizing \textit{multi}-level information can potentially outperform those using single-level information. Yet, there have not been many efforts to utilize such information.

\begin{figure}[t!]
    \centerline{
    \includegraphics[width=0.99\columnwidth]{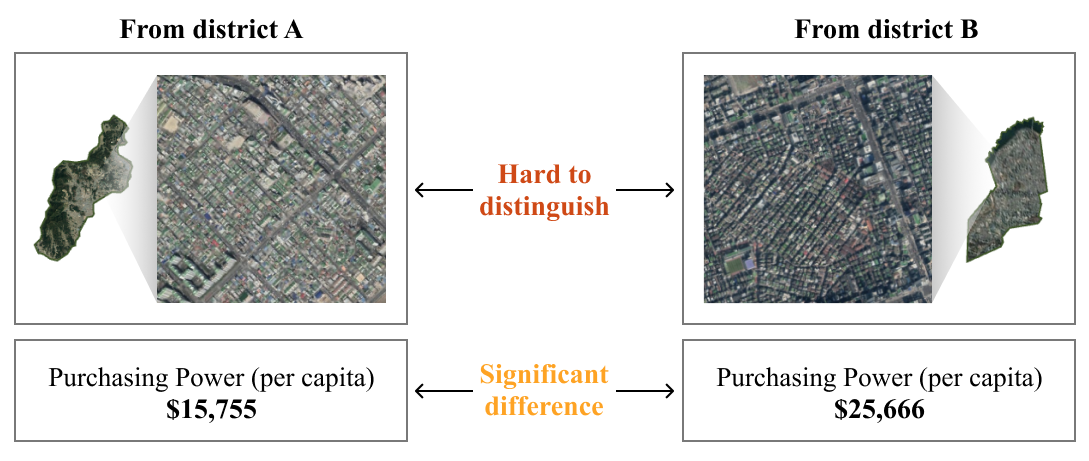}}
    \caption{A motivational example shows that even though satellite images from two urban regions do not display noticeable differences, their economic indicators show a large gap.
    (A: a district in Daejeon, 
    B: Gangnam district in Seoul, South Korea)
    }
    \label{fig:intro}
\end{figure}

Figure~\ref{fig:intro} shows an example where two urban grids exhibit varying degrees of purchasing power despite their visual similarity. Both images appear highly urbanized when compared as cropped-out grid tiles. However, a large economic gap exists at the hyperlocal level because they belong to districts of substantially different economic scales. This discrepancy is due to the presence of an agglomeration effect or the productivity benefit arising in clustered urban communities with dense populations and industries~\cite{marshal1890, durantonetal2004}. As studies have shown~\cite{hui2020predicting}, economies of scale are common in the urban economy; yet this information cannot be observed easily from a single grid image. Instead, understanding the interconnection of hyperlocal economies that make up a district can help estimate economic indicators.

\begin{figure*}[t!]
    \centering
    \includegraphics[width=1.99\columnwidth]{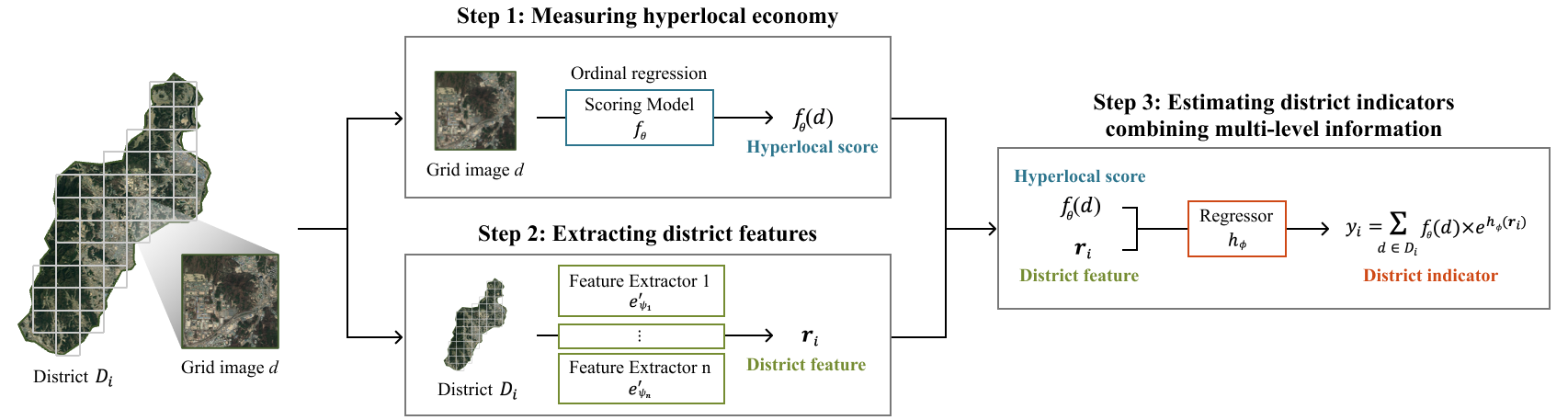}
    \caption{Illustration of the model architecture. Images collected from each district $D_i$ are put into our models to calculate the hyperlocal scores and district scaling factor. These outputs are combined to predict the economic indicator $y_i$ of the given district $D_i$.}
    \label{fig:model_summary}
\end{figure*}

We present a novel method that utilizes \textit{multiple} levels of geographic information to predict economic indicators accurately. The model first measures the hyperlocal economy by inferring the relative degree of economic development for individual grid images via ordinal regression. Next, the interconnected relationship among small grid areas within the same district is summarized as district features. These features act as a scaling factor across districts to differentiate the scores of otherwise similar-looking grids. The model finally estimates economic indicators of districts by jointly utilizing the hyperlocal-level predictions and district-level scaling factors.

Experiments based on a million-scale data suggest that the district scaling factor, which captures the power-law relationship of urban economies (Figure~\ref{fig:pareto}), contributes to the substantial performance gain of the proposed model in predicting key economic indicators compared to extant baselines that rely on single-level information (Table~\ref{tab:baseline_results} and Figure~\ref{fig:bar_graph}).
In addition, experiments show that our model is robust against scarce data conditions (Figure~\ref{fig:wo_augmentation_entire}), a quality that is critical for practical policy implications. We present economic interpretations of the model predictions based on Zipf’s law and discuss implications for measuring economic inequality in underdeveloped countries. We release our code via  GitHub.\footnote{github.com/deu30303/DistrictEffect}

\section{Related Work}
Satellite imagery conveys visual information about the earth's surface and has shown widespread usage in economic predictions, replacing labor-intensive survey data sources~\cite{albert2017using}. Early methods used nighttime imagery as a proxy of economic activity~\cite{bagan2015analysis,ghosh2013using}. Nowadays, high-resolution daytime images are in use.

Prior to satellite imagery, the ground truth of prediction models was mainly the household surveys collected on small regions; they report hyperlocal indicators such as wealth index or public health~\cite{jean2016combining,tingzon2019mapping, bondimapping}. The prediction models for those indicators utilize not only the satellite imagery but also the road information from OpenStreetMap, which is the editable map that contains geodata across the world that is actively supporting user (human) intervention~\cite{haklay2008openstreetmap}. When the human-mediated data is applied, the latest model also states that a hyperlocal unit's economy can be analyzed even without the ground truth labels. The model can be trained via a human-guided comparison in a weakly supervised fashion~\cite{han2020learning}.

Recent studies have utilized both surveys and daytime satellite images on districts. Embedded statistics are used to summarize feature vectors of a target district. For example, READ first applied principal component analysis (PCA) for reducing dimensions of each image's representation and summarized each district into a fixed-length vector of multiple descriptive statistics~\cite{han2020lightweight}. These fixed numbers of variables enable a regression to predict the district economy. Daytime images have also been used for quantifying the level of urban green space~\cite{kwon2021epjds}.
\section{Model}
Assume multiple districts of arbitrary shape and size compose the territory of a country. We denote a set of grid satellite images in the $i$-th district as $\district_{i} = \{\img_1,...,\img_{n_i}\}$. Our task is to estimate the economic indicator $y_i$ of district $i$, given $\district_{i}$. We propose a model that measures the economy at two levels: (1) a hyperlocal score $f_\theta(\img)$ and (2) per district scaling factor $g_\varphi(\district)$. While the former represents the degree of economic development of the hyperlocal region captured by each satellite image, the latter represents the aggregated effect among multiple hyperlocal regions that compose a district. These two elements are then jointly used to predict the district economy as $y_i=\sum_{\img\in \district_i} f_\theta(\img) \cdot e^{g_\varphi(\district_i)}$. Figure~\ref{fig:model_summary} illustrates the model architecture.

End-to-end training on district-level supervision can lead to overfitting since the number of the annotated districts is typically much smaller than the satellite image count. To alleviate this problem, we propose a stage-wise training using different datasets and training signals as follows:
\begin{itemize}
    \item \textbf{Step 1.} Given a small dataset with proxy labels relevant to our target economic variables, train the hyperlocal score model $f_\theta$ under an ordinal regression objective.
    
    \item \textbf{Step 2.} 
    Learn a district-level representation by training an image-level encoder $e_\psi$ in a semi-supervised manner, then aggregate them into a fixed-size representation ${\bf r}_i$.
    
    \item \textbf{Step 3.} 
    Given the hyperlocal score model $f_\theta$ and district-level representation ${\bf r}_i$, learn a regression model $h_\phi$ as $g_\varphi(\district_i) = h_\phi(\mathbf{r}_i)$ using the district-level supervision.
\end{itemize}

\subsection{Step 1. Measuring Hyperlocal Economy}

Let $\img_{n}$ be the $n$-th satellite image in the entire image set of a country. Then the first step aims to learn a score model $f_\theta$ that predicts the degree of economic development $s_{n}$ of a satellite image $\img_{n}$ (i.e., $s_{n} = f_{\theta}(\img_{n})$). Economic statistics are typically available at the district level ($y_i$) rather than for each satellite image. We instead employ an auxiliary dataset $\mathcal{X}$ with surrogate labels indicating the coarse category of an image as weak supervision. We choose a random set of 1,000 images and manually labeled them as urban, rural, or uninhabited (i.e., $\mathcal{X} = \{(\img_n, \hat{\mathbf{y}}_n); n \in (1, ..., 1000)\}$).\footnote{Each data was annotated by four annotators, whose aggregated decisions were used as soft labels. Fleiss kappa, which assesses the degree of agreement, was 0.72.}

Since we want our model to produce a continuous score from the (soft) categorical label, we employ ordinal regression with the linear property. Ordinal regression establishes an ordinal relationship between classes and handles ordered target variables~\cite{brant1990assessing}. We aim to preserve the order of economic development among three surrogate labels as follow:
\begin{linenomath}
\begin{align}
\text{class of $\img$} = 
\left\{
\begin{array}{ll}
     \text{uninhabited} & \mbox{if } f(\img) < t_1 \\
     \text{rural} & \mbox{if } t_1 \leq f(\img) < t_2   \\
     \text{urban} & \mbox{if } t_2 \leq f(\img)
\end{array}
\right.
\end{align}
\end{linenomath}
where $f$ is an abbreviation of $f_{\theta}$ and $t_1$, $t_2$ are thresholds for distinguishing two adjacent classes ($t_1 < t_2$). Then, a logit vector is used to train the score model:
\begin{linenomath}
\begin{align}
    l(\img) = [t_1 - f(\img),\ \min (f(\img) - t_1, t_2 - f(\img)),\ f(\img) - t_2].
\label{eq:logit}
\end{align}

\end{linenomath}

Eq.~\eqref{eq:logit} ensures that each element in the logit vector represents the distance between score and threshold. When the score grows over $t_2$ (i.e., urban), the third component's value in $l(\img)$ becomes positive, while the remaining becomes negative. The second component becomes positive if the score lies between the two thresholds (i.e., rural). Therefore, applying a softmax function to this logit $l(\img)$ and performing classification guarantees that the score model $f_\theta$ preserves the predefined label order.
Unlike the existing approaches in ordinal regression employing the non-linear output kernels~\cite{liu2018constrained,brant1990assessing}, the linearity in Eq.~\eqref{eq:logit} encourages our model to produce continuous scores. This property is important to learning the continuous economic scale of an image based on discrete surrogate labels. We clamp the scores into the fixed range [$t_{min}$, $t_{max}$] and train our model to minimize the cross-entropy loss for numerical stability:
\begin{linenomath}
\begin{align}
\mathcal{L}_{class} =  - \frac{1}{|\mathcal{X}|} \sum_{(\img,\hat{\mathbf{y}}) \in \mathcal{X}}{\hat{\mathbf{y}}^\mathrm{T} \cdot\text{LogSoftmax}(l(\img))} \label{eq:loss_class}
\end{align}
\end{linenomath}

\begin{figure}[t!]
    \centerline{
    \includegraphics[width=0.8\columnwidth]{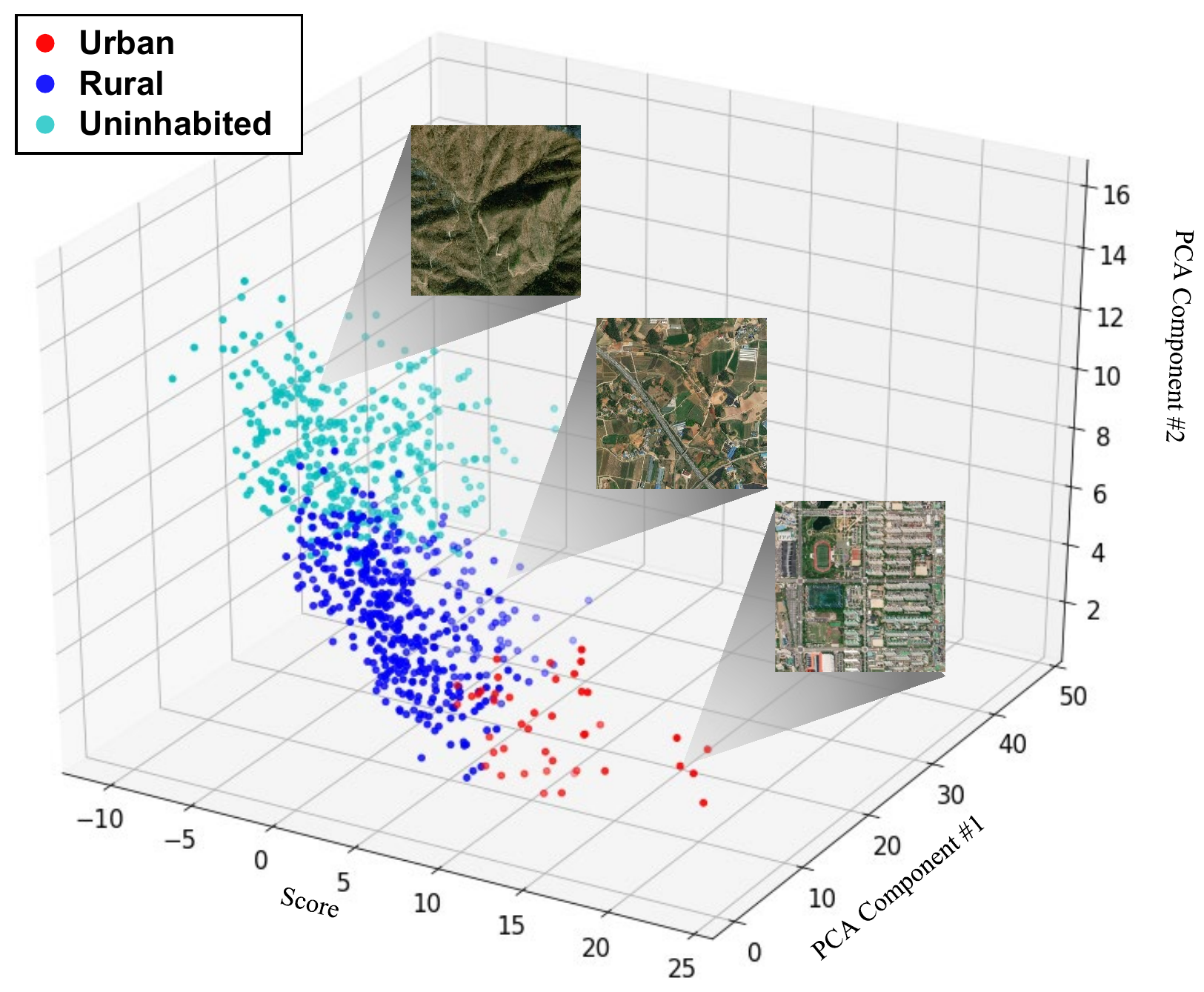}}
    \caption{
    Visualization of score distribution obtained by our model (\emph{i.e.}, $s=f_\theta(\img)$). The model clearly separates images in different classes while providing relative orders among in-class images.
    }
    \label{fig:orders}
\end{figure}

We use the outputs from the model $s=f_\theta(\img)$ as an economic scale of the hyperlocal image $\img$. Figure~\ref{fig:orders} illustrates the score distribution in the projected space obtained by PCA. The learned model separates different classes well while ordering images even within the same class. In the experiment, we verify that such relative order of images predicted by our score model is aligned well with actual economic development among regions.

\subsection{Step 2. Extracting District Features}
Although the score model provides a relative economic score on a hyperlocal image, predicting district-level economy requires understanding the interconnection between hyperlocal images within the district. For example, when urban towns are geographically close, an agglomeration effect allows for a much larger economic development than when urban towns are positioned far from one another~\cite{durantonetal2004}. Step 2 aims to encode such information into the district-level representation.

\subsubsection{Extracting features from an image}

Before incorporating district features, we first train an encoder $e_\psi$ as a feature extractor for each satellite image.
Our encoder training is guided by the surrogate loss in Eq.~\eqref{eq:loss_class} but also regularized by additional clustering loss to improve the generalization and enhance the correlation between visually similar images~\cite{han2020mitigating}. We employ a large set of unlabeled satellite images $\mathcal{U}$ to compute the regularization\footnote{We eliminate the uninhabited images in $\mathcal{U}$ by applying the score model $f_\theta$, as they are irrelevant to economic measures.}
and adopt DeepCluster for clustering ~\cite{caron2018deep}. 
By computing the k-means algorithm over the embedded vectors, DeepCluster generates clusters that can be regarded as pseudo-label. These pseudo-labels are then utilized to update the encoder by minimizing the following classification loss: 

\begin{linenomath}
\begin{align}
\mathcal{L}_{cluster} = \frac{1}{|\mathcal{U}|} \sum_{\img \in \mathcal{U}} \bar{\mathbf{y}}^\mathrm{T} \cdot\text{LogSoftmax}(W\cdot e_\psi(\img)),\label{eq:deepcluster}
\end{align}
\end{linenomath}

where $\bar{\mathbf{y}}$ denotes the pseudo-label of the image from k-means clustering, and $W$ is a weight matrix that projects the encoder output to the pseudo-label space. (\emph{i.e.}, cluster head) \looseness=-1

We repeated the clustering algorithm for urban (i.e., $f_{\theta}(\img) \geq t_2$) and rural classes (i.e., $f_{\theta}(\img) < t_2$) respectively and got an equal number of $n_c$ clusters from both classes. Finally, the encoder is trained in a semi-supervised way with multi-task loss using two datasets $\mathcal{X}$ and $\mathcal{U}$ to preserve the label order while gathering similar visual traits in embedding space (Eq.~\ref{eq:total_loss}). 
\begin{linenomath}
\begin{align}
\mathcal{L}_{encoder} = \mathcal{L}_{class} + \lambda \times \mathcal{L}_{cluster} \label{eq:total_loss}
\end{align}
\end{linenomath}
We initialize the encoder $e_\psi$ using the score model's parameters $f_\theta$ except the last layer. We observed that the adjusting parameter $\lambda=1$ works well in all experiment settings.

\subsubsection{Summarizing district features}
Given the encoder $e_\psi$, the next task is to aggregate the image-level features to obtain the district-level representation. Districts can be of any shape and size, and the number of images for districts can show orders of magnitude differences. Accordingly, the feature set's size extracted from satellite images in each district may vary. Hence, summarizing a different number of satellite images' features into a fixed-length representation is essential to handle the district scaling factors.

Embedded spatial statistics can summarize all feature vectors within a district. This approach extracts fixed-length representations from different numbers of images. We applied PCA to reduce the dimension of the encoder representation\footnote{We set the number of principal components to three, as they explain about 80\% of the total variance.} and aggregated all representations within the district and calculated the descriptive statistics, such as mean and standard deviation, to produce a fixed-length vector. We also add the summarization of the hyperlocal scores $f_\theta$ within the district to provide hyperlocal information. The final district feature vector $\mathbf{r}_i$ is defined as follows:
\begin{linenomath}
\begin{align}
\mathbf{r}_{i} = [ \mu(e'_\psi(\district_i)), \sigma(e'_\psi(\district_i)), \Sigma_{\img\in D_i}f_\theta(\img) ],
\end{align} 
\end{linenomath}
where $\mu(\cdot)$ and $\sigma(\cdot)$ denote mean and standard deviation, respectively, and $e'_\psi$ denotes the projected encoder outputs by the PCA. We calculate $\mathbf{r}_i$  for every district and utilize it as an input to the regressor $h_\phi$, which estimates the district scaling factor.

\subsubsection{Ensemble approach for better representation.}
An ensemble approach is considered to improve the representation quality. This method trains several models on the same objective from different conditions and is robust against uncertain quantitative predictions~\cite{opitz1999popular, ovadia2019can, park2021improving}. 

We enrich the representation ${\bf r}$ by the ensemble of encoders $\{e'_{\psi_2}, ..,, e'_{\psi_M}\}$ as follows:
\begin{linenomath}
\begin{align}
\boldsymbol{\mu}_{i} &= [\mu(e'_{\psi_1}(\district_i)),..., \mu(e'_{\psi_M}(\district_i))] \\
\boldsymbol{\sigma}_{i} &= [ \sigma(e'_{\psi_1}(\district_i)),..., \sigma(e'_{\psi_M}(\district_i))] \\
\mathbf{r}_{i} &=  [ \boldsymbol{\mu}_{i}, \boldsymbol{\sigma}_{i}, \Sigma_{\img\in D_i}f_\theta(\img)  ]
\end{align}
\label{eq:ensemble}
\end{linenomath}
where $[\cdot]$ denotes the concatenation.

To introduce the diversity to the ensemble of encoders, we employ two strategies: (1) training encoders using different data sources or (2) randomizing cluster assignment in DeepCluster. As an additional data source for ensemble, nightlight intensity data is used as a proxy to train the extra feature extractor. The model can learn about the general characteristics of urban development by maximizing the Pearson correlation between nightlight intensity and hyperlocal scores from entire daytime satellite images of a country. To introduce randomization during clustering, we trained multiple encoders by optimizing Eq.~\eqref{eq:deepcluster} with a various number of clusters $n_c\in\{0, 30, 90\}$ in two data sources (i.e., $\mathcal{X}$ and nightlight intensity) independently. Thus, six models are implemented to construct the final representation vector $\mathbf{r}_i$.

\subsection{Step 3. Estimating District Indicators Combining Multi-Level Information}
Given the hyperlocal score model $f_\theta$ and district-level representation ${\bf r}_i$, the last step combines the two to estimate the economic development of districts.
The prediction on $i$-th district-level economy is defined as follow:
\begin{linenomath}
\begin{align}
y_i = \sum_{\img \in \district_i}{f_{\theta}(\img)} \times e^{h_{\phi}(\mathbf{r}_i)}.
\label{eq:scaling}
\end{align}
\end{linenomath}
The hyperlocal score model is trained independently of the target district economic indicators. In contrast, the district scaling model is trained to adjust the prediction scores according to the target economic indicators. 
We train the regression model $h_\phi$ in Eq.~\eqref{eq:scaling} using the supervision on the district-level economic measure as follow:
\begin{linenomath}
\begin{align}
h_{\phi}(\mathbf{r}_i) \approx \ln{y_i \over \sum_{\img \in \district_i}{f_{\theta}(\img)}}\ \  \label{eq:regressor}
\end{align}
\end{linenomath}

\begin{table*}[t]
\centering

\resizebox{2.0\columnwidth}{!}{
\begin{tabular}{l|c c c c c}
\toprule
\multicolumn{1}{c|}{\textbf{Model}} & \multicolumn{1}{c}{\textbf{Total population}} & \multicolumn{1}{c}{\textbf{Purchasing power}} &
\multicolumn{2}{c}{\textbf{Energy consumption}} &
\multicolumn{1}{c}{\textbf{GRDP}} \\
&&& Domestic & Total & \\
\midrule 
Nightlight Proxy &  0.1144 $\pm$ 0.3505  &  0.0999 $\pm$ 0.3401 & 0.1484 $\pm$ 0.3187 & -0.9765 $\pm$ 1.0647 & -0.9932 $\pm$ 1.0386 \\
Tile2Vec &   0.2320 $\pm$ 0.1225  &  0.2199 $\pm$ 0.1547 & 0.3196 $\pm$0.1580 & -0.1572 $\pm$ 0.4397 &  -0.3217 $\pm$ 0.3237 \\
SimCLR &   0.4081 $\pm$ 0.1217  &  0.4271 $\pm$ 0.1342 & 0.4093 $\pm$0.1291 & 0.0319 $\pm$ 0.2619 &  -0.2675 $\pm$ 0.2682 \\
READ &   0.5920 $\pm$ 0.0979  &  0.5286 $\pm$ 0.1052 & 0.5632 $\pm$0.1030 & 0.2412 $\pm$ 0.1917 &  0.2036 $\pm$ 0.2273 \\
\midrule
Full model (ours)& \textbf{0.8149} $\pm$ 0.0721  & \textbf{0.8212} $\pm$ 0.0721 & \textbf{0.8553} $\pm$ 0.0447 & \textbf{0.4853} $\pm$ 0.1923 & \textbf{0.4568} $\pm$ 0.1701 \\
Without ensemble & 0.7296 $\pm$ 0.1262 & 0.7152 $\pm$ 0.1422 & 0.7721 $\pm$ 0.1033 & 0.3775 $\pm$ 0.2388 & 0.2968 $\pm$ 0.2102\\
Without fine-tuning & 0.7781 $\pm$ 0.1067  &  0.7790 $\pm$ 0.1009 &  0.8266 $\pm$ 0.0633 & 0.4783 $\pm$ 0.2013 & 0.4023 $\pm$ 0.2013\\
Without hyperlocal scoring  & 0.5249 $\pm$ 0.1045 & 0.5058 $\pm$ 0.0812 & 0.4624 $\pm$ 0.1072 & 0.2088 $\pm$ 0.1531 & 0.2011 $\pm$ 0.2435 \\
\bottomrule
\end{tabular}}
\caption{Evaluation results including four existing baselines and ablation studies. Performances are evaluated on four representative economic indicators: total population, purchasing power, energy consumption, and gross regional domestic product (GRDP) of districts in South Korea. Every ground-truth value is in its original scale, and R-squared values are reported for evaluation.}
\label{tab:baseline_results}
\end{table*}

\subsubsection{District data augmentation.}

Although the optimization of Eq.\eqref{eq:regressor} is limited with only respect to the regressor $h_\phi$, the amount of training data with district-level economy labels is still insufficient to avoid overfitting. To alleviate this challenge, we present a new data augmentation technique. Let $N$ be the number of annotated districts with the economic label. Then we generate a new larger district by randomly selecting two districts and unifying them, i.e., $\district_{i\cup j} \equiv \district_i \cup \district_j$. Accordingly, we assign an economic label to these new districts by aggregating their respective labels by summation (or weighted average). The proposed data augmentation combinatorially increases the number of training instances (i.e., $N + {N \choose 2}$), thereby preventing the overfitting. 

\section{Training Setups}
\subsection{Datasets}
\subsubsection{Satellite images} 
We use high-resolution daytime satellite images of five Asian countries: South Korea, Malaysia, Vietnam, the Philippines, and Thailand. We chose Korea for its rich datasets and ground truth and the others for predicting different degrees of inequality in developing countries. A GIS software company, Esri, provides daytime satellite image tiles at various zoom levels (Z). The zoom level determines a tile size, where $Z=k$ indicates that the tile's width/height covers over $2^k$ of the entire earth's longitude/latitude. For example, an image covering the whole earth has a zoom level of 0, and both the width and height become half at each zoom level increment. Since the number of pixels per image tile remains identical (256 by 256), a higher zoom level indicates a richer resolution. We utilized images of $Z=15$, which has $4.7m$ per pixel resolution for all experiments. Images are from 2017 to 2019.  

\subsubsection{Economic indicators}
We use four ground truth indicators that explain the district-level economy: total population, purchasing power (i.e., the disposable income after taxes and transfers), energy consumption, and gross regional domestic product (GRDP). Esri Demographics makes the first two economic indicators available for 135 countries via its ArcGIS GeoEnrichment API.\footnote{Data from Michael Bauer Research GmBH. More details are listed on doc.arcgis.com/en/esri-demographics/data} For the latter two, we use the official statistics released by the studied country. 

\subsection{Training Details}
\subsubsection{Hyperlocal score model} $f_\theta$ in step 1 was trained for 100 epochs with batch size 50. Thresholds for distinguishing two adjacent classes (i.e., urban-rural and rural-uninhabited) were set to 0 and 10. For numerical stability, we clamped the scores into the fixed range [-10, 20]. Adam optimizer with a learning rate 1e-4 was utilized.  

 \subsubsection{Feature extractor model} $e_\psi$ in step 2 utilized multi-task loss from two datasets: $\mathcal{X}$ and $\mathcal{U}$. Batch sizes for $\mathcal{X}$ and $\mathcal{U}$ were set to 40 and 256, respectively. The model was trained for 5 epochs. For ensemble, six extractors $\{e'_{\psi_1}, e'_{\psi_2}, ..,, e'_{\psi_6}\}$ were implemented by altering the number of clusters $n_c\in\{0, 30, 90\}$, and by introducing two different data sources (i.e., $\mathcal{X}$ and nightlight intensity) independently.  In the case of the nightlight intensity dataset $\mathcal{N}$, the encoder model was trained to maximize the Pearson correlation between nightlight intensity $\Tilde{y}$ and hyperlocal scores from entire daytime satellite images of a country. ResNet-18 was used as a backbone network for both hyperlocal score model $f_\theta$ and feature extractor $e_\psi$.
 
 \subsubsection{Final regressor} $h_\phi$ in step 3 is fit with ground-truth economic indicators $y$. A random forest regressor with 200 estimator trees was used.

\section{Experiments}
\subsection{Performance Evaluation}
\subsubsection{District-level evaluation} The first evaluation uses multiple district-level ground-truth statistics. We randomly split the data into a training set and a test set with an 80\%-20\% ratio 100 times. We report the R-squared values as an evaluation metric to show how well our predictions approximate the ground truth; this metric shows the proportion of the variance for a dependent variable explained by a regression model. Previous work on economic predictions~\cite{jean2016combining,han2020lightweight} employed log-scaled ground truth values for evaluation. However, we evaluate all models with their original scales for practical usage.

We use seven baselines: four from existing studies and three from ablations. All baselines had an identical setting of the split ratio and the backbone network. 
\textbf{(1) Nightlight Proxy} uses the nightlight intensity as a proxy to train an encoder~\cite{jean2016combining}. It extracts features from daytime images with an encoder, and features are summarized to predict economic indicators by averaging.
\textbf{(2) Tile2Vec} adopts a triplet loss using geospatial distance information to represent individual satellite images in an unsupervised manner~\cite{jean2019tile2vec}.
 We summarize district features from the learned model by averaging. 
\textbf{(3) SimCLR} is a self-supervised framework of visual representations~\cite{chen2020simple}. It learns visual features by maximizing agreement between different views of the same image. The learned features are summarized by averaging and eventually used for prediction. This model was originally proposed for satellite imagery, but we apply it for its superior representation quality.
\textbf{(4) READ} is the weakly supervised model that can summarize visual features from districts to estimate population density~\cite{han2020lightweight}. Since this model predicts population density, we multiply the predicted number by the area size to report the total population. 
The next are ablations, where we vary the proposed model by intentionally missing a key component. 
\textbf{(5) Without ensemble} is an ablation of our model without the ensemble approach Eq~7--9.
\textbf{(6) Without fine-tuning} is an ablation without the fine-tuning step 
Eq~4--5.
\textbf{(7) Without hyperlocal scoring} is an ablation that only uses district scaling factors. 

\begin{figure}[t!]
\centering
    \begin{subfigure}[b]{0.23\textwidth}
      \includegraphics[width=\textwidth]{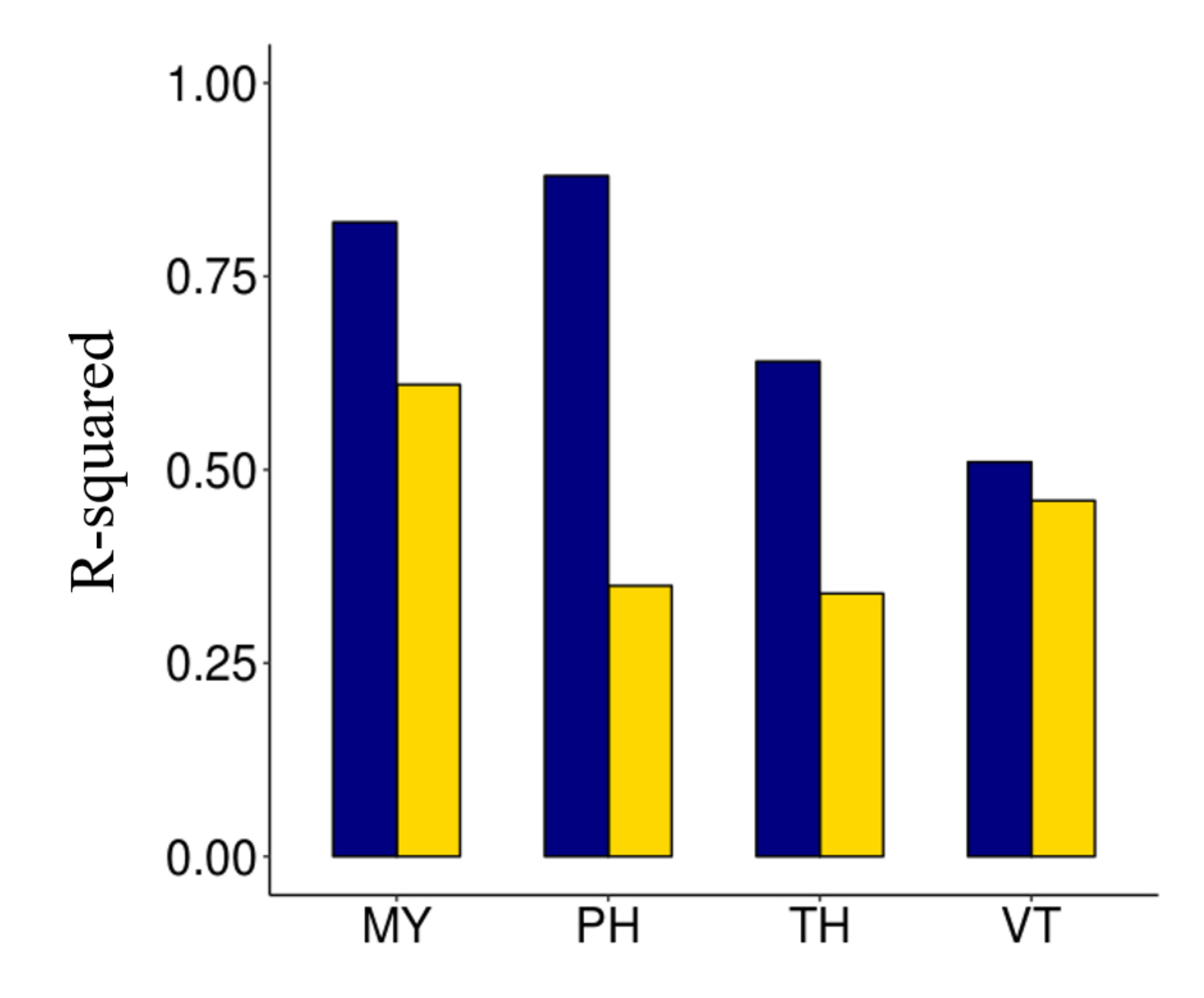}
      \caption{Total population}
    \end{subfigure}
    \begin{subfigure}[b]{0.23\textwidth}
      \includegraphics[width=\textwidth]{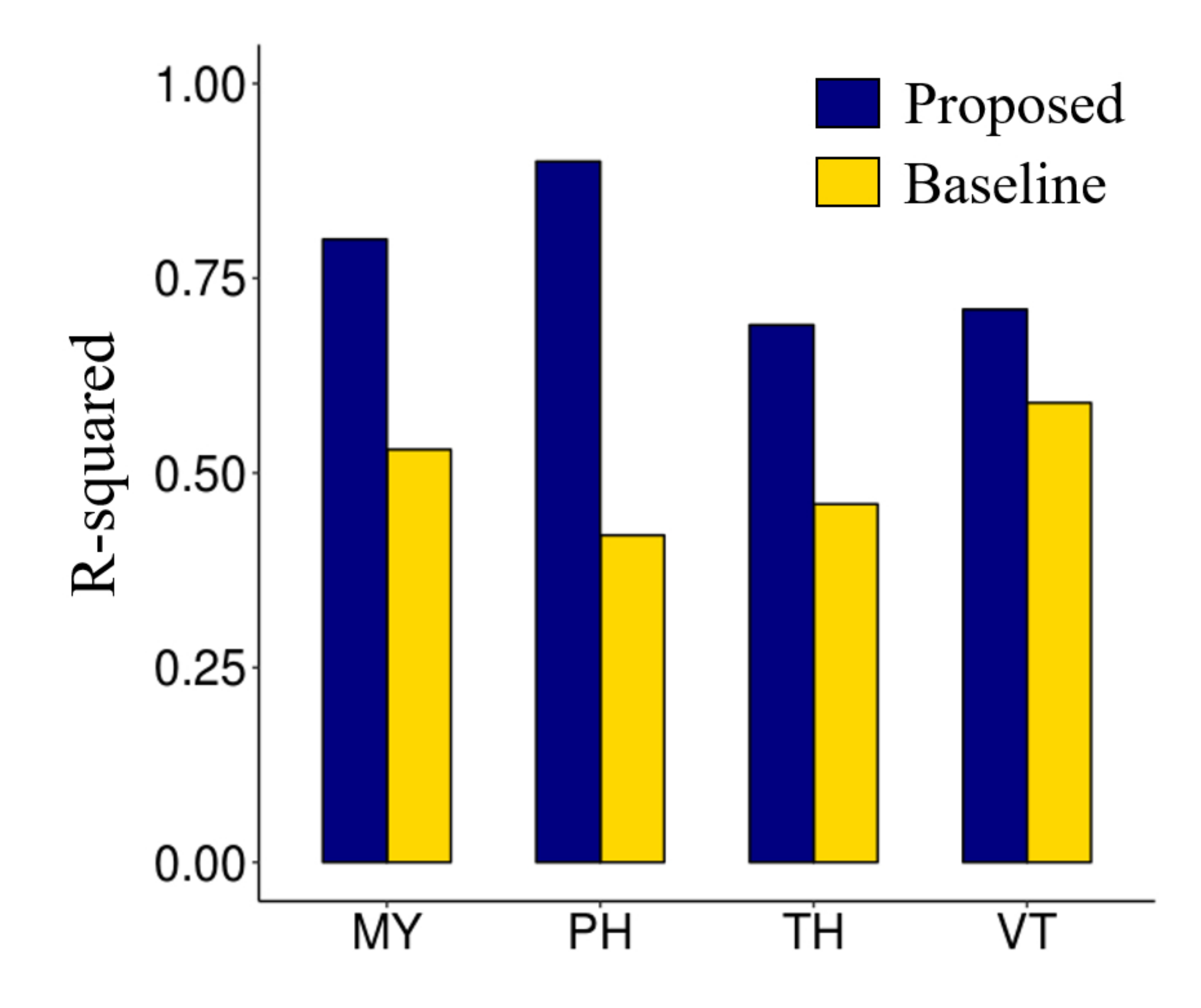}
      \caption{Purchasing power}
    \end{subfigure}  
\caption{Prediction results over four Asian countries on two economic indicators: total population and purchasing power. The R-squared values are compared for the proposed model and the best performing baseline. MY is Malaysia, PH is the Philippines, TH is Thailand, and VT is Vietnam.}
\label{fig:bar_graph}
\end{figure}

Table~\ref{tab:baseline_results} shows that our model outperforms all baselines by a large margin across all tasks, for the case of predictions on South Korean data. The Nightlight Proxy model shows relatively poor performance than the other daytime image models due to the light saturation, nighttime glare, and missing detailed landscape information. The light saturation is known to occur when the light intensity above a certain level does not increase, making it challenging to distinguish subtle differences across urban districts~\cite{zheng2018method}.
The table also shows that removing any key component in the proposed model lowers the performance, indicating all three components contribute to predicting economic indicators. Among them, the hyperlocal score has the most critical role.

The proposed model also performs well in the other four Asian economies. We train the model independently for each country and test the performance with an 80-20 train-test split. All results are intra-national predictions. Figure~\ref{fig:bar_graph} compares the prediction result against the best performing baseline, READ~\cite{han2020lightweight}. The proposed model performs substantially better than the baseline for all countries studied.

\subsubsection{Hyperlocal-level evaluation}
Given the importance of the hyperlocal scoring, we also compare two variants of the model for the grid-level prediction: One is the original scores $f_\theta(\img)$ and the other is the adjusted scores $f_\theta(\img) \cdot e^{g_\varphi(\district)}$ that are re-scaled by the district scaling factor. The hyperlocal economy can be computed by aggregation of a geo-located economic indicator within the arbitrary area. Gross floor area (GFA), the amount of space filled by a building, could work as an excellent ground-truth indicator for aggregation, publicly released by the Ministry of Land, Infrastructure, and Transport from South Korea. 

Table~\ref{table:local_eval} shows that while both hyperlocal scores distinctly correlate with ground-truth GFA, the adjusted hyperlocal score exhibits more precise prediction. The original hyperlocal scores $f_{\theta}(\img)$ showed a moderate correlation (Pearson correlation = 0.626). The adjusted hyperlocal score exhibits a more precise prediction (Pearson correlation = 0.777). These findings underscore the importance of multi-level considerations.

\begin{table}[t!]
\centering
\resizebox{0.99\columnwidth}{!}{
\begin{tabular}{l|cc}
\toprule
Correlation & Original ($f_\theta(\img)$) & Adjusted ($f_\theta(\img) \cdot e^{g_\varphi(\district)}$) \\ 
\midrule
  Pearson   & 0.626  &     \textbf{0.777}  \\ 
  Spearman  & 0.788 &     \textbf{0.790} \\   
\bottomrule
\end{tabular}}
\caption{Correlation between the gross floor area (GFA) and hyperlocal scores measured in two manners.}
\label{table:local_eval}
\end{table}

\begin{figure}[t!]
    \centerline{
    \includegraphics[width=0.8\columnwidth]{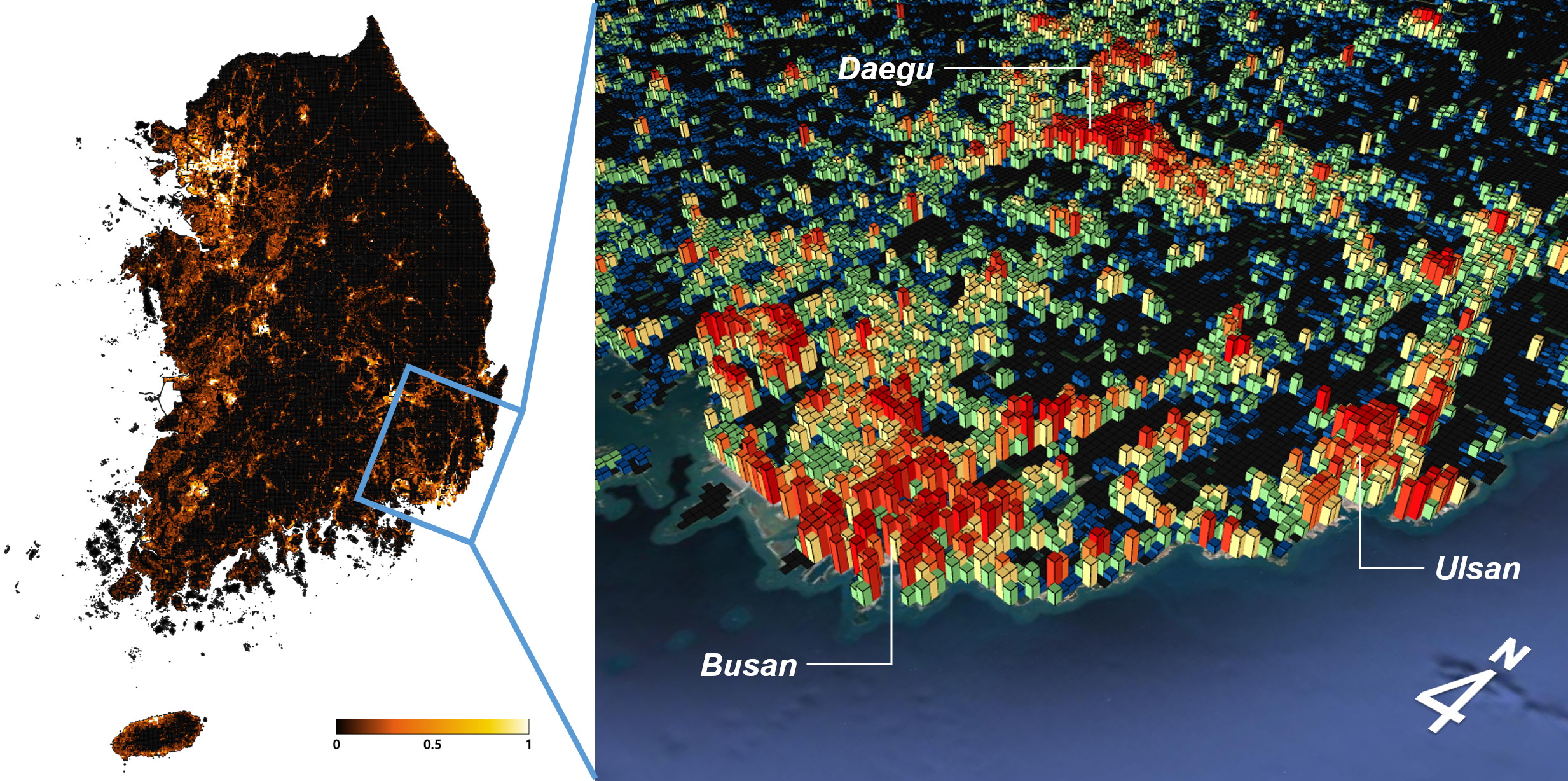}}
    \caption{A 3D visualization of the economic development predicted for each hyperlocal area by the model for South Korea (left) and its Busan metropolitan area (right).}
    \label{fig:korea_vis}
\end{figure}

\begin{figure*}[t!]
\centering
    \includegraphics[width=1.94\columnwidth]{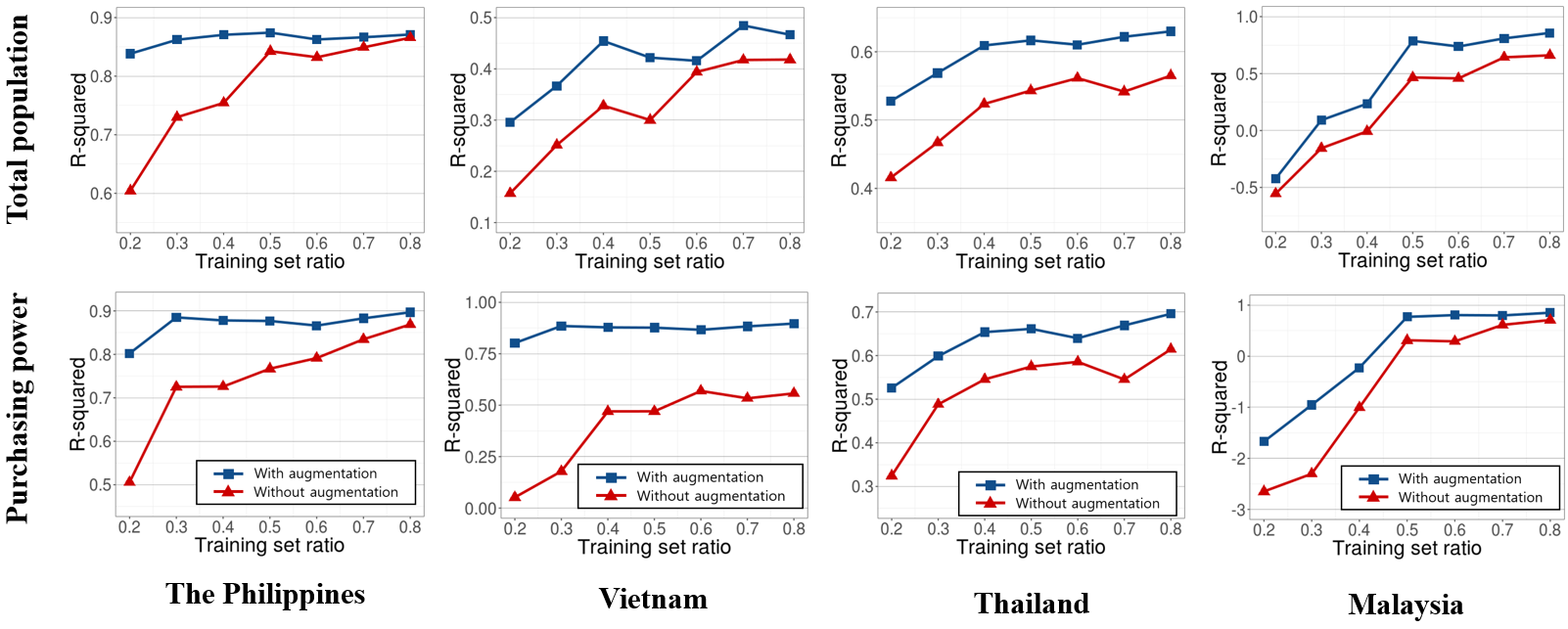}
\caption{The effect of the training-to-testing ratio on prediction performance for four countries data: the Philippines, Vietnam, Thailand, and Malaysia. District data augmentation keeps the model robust under a scarce training set setting.}
\label{fig:wo_augmentation_entire}
\end{figure*}

Figure~\ref{fig:korea_vis} visualizes the hyperlocal economic development score predicted by the proposed model over South Korea. The zoom-in image on the right shows the development level predicted over the hyperlocal regions (indicated by the height in a 3D plot), demonstrating the model's ability to learn economic indicators over \textit{multiple} geographic levels.

\subsection{Robustness Testing} 
Collecting up-to-date economic information requires a considerable amount of resources. Many underdeveloped and developing economies suffer from such data deficiency problems. Here we test two practical scenarios to deal with limited data: (1) utilize a smaller subset of the training dataset and (2) transfer the model parameters learned from neighboring countries. Additionally, we also study (3) the effect of the backbone network choice for empirical usage.

\begin{figure}[t!]
\centering
    \begin{subfigure}[b]{0.23\textwidth}
      \includegraphics[width=\textwidth]{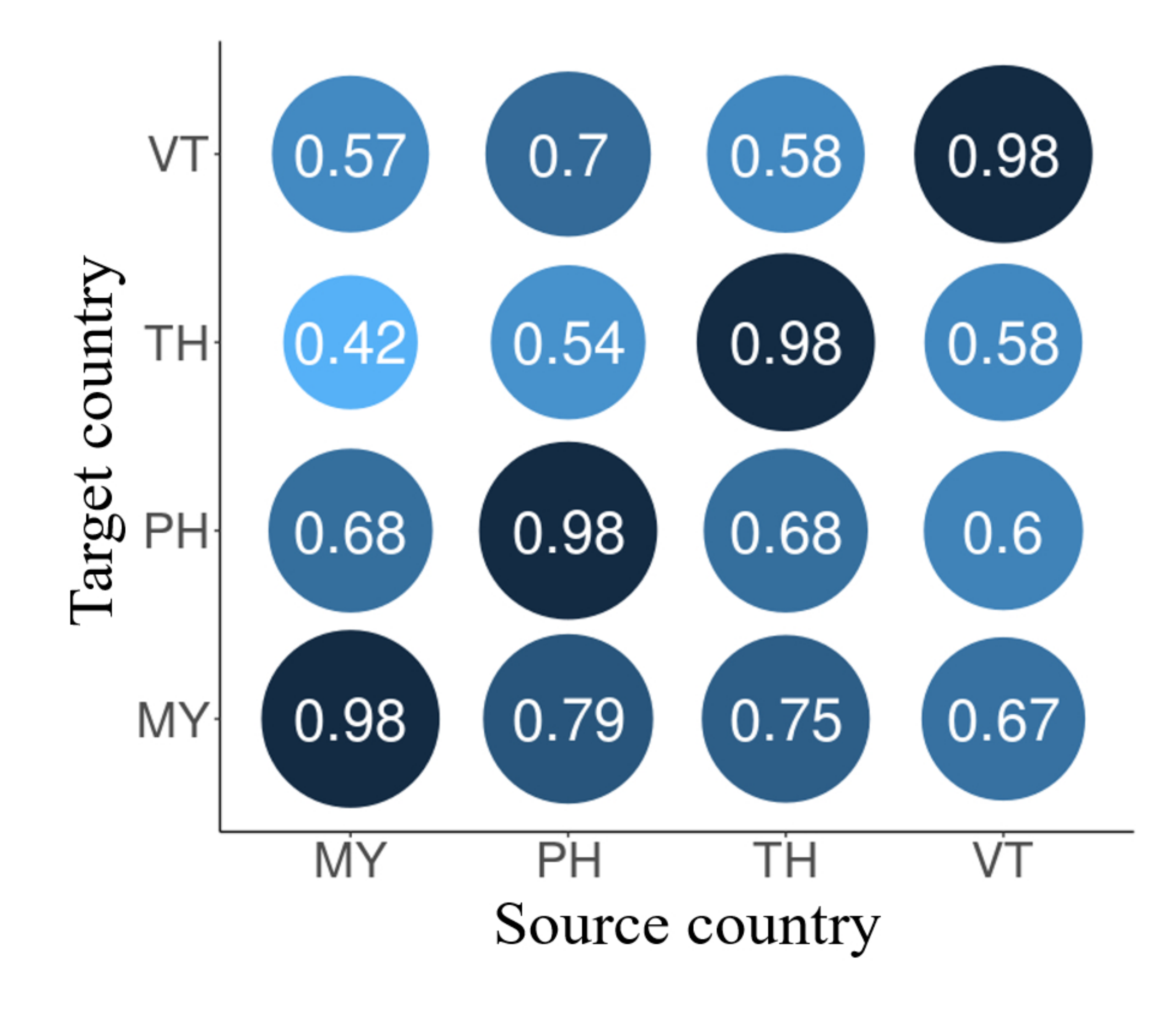}
      \caption{Total population}
    \end{subfigure}
    \begin{subfigure}[b]{0.23\textwidth}
      \includegraphics[width=\textwidth]{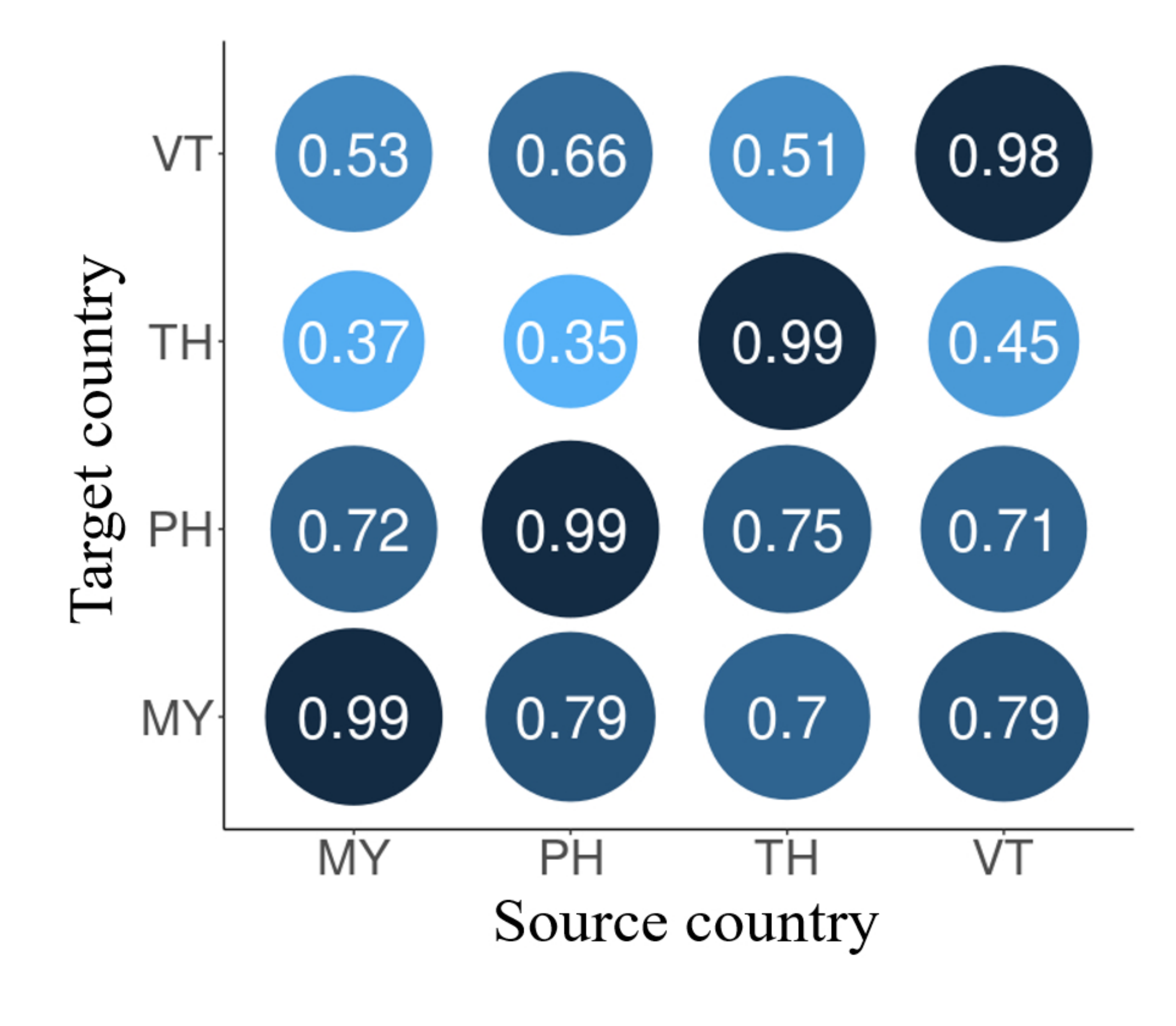}
      \caption{Purchasing power}
    \end{subfigure}  
\caption{The Spearman correlation for the transferability test on two economic indicators: total population and purchasing power. The model is trained on one source country and then evaluated on other target countries without fine-tuning.}
\label{fig:cross_country}
\end{figure}

\subsubsection{Dataset sizes} We first investigate how well the model can learn from limited data Here we varied the ratio of training and testing from 0.2 to 0.8 then iteratively computed the R-squared values of predictions over four Asian countries: the Philippines, Vietnam, Thailand and Malaysia.
Figure~\ref{fig:wo_augmentation_entire} shows that the proposed model is robust against data shortage. This superiority originates from the district data augmentation method explained in the model section, which was intended to prevent overfitting. Results suggest that the district data augmentation technique is also indispensable against data scarcity. When the model does not incorporate augmentation, prediction quality drops for small training data.

\subsubsection{Transferability} The transferability test checks whether model parameters learned from one country (i.e., source) can be used in another country (i.e., target). The source-data trained model is used to predict a target country's district-level economy. We varied the source and target pairs and computed the performance using the Spearman correlation of the predicted economies and the ground truth statistics. The diagonal line in Figure~\ref{fig:cross_country} shows the highest correlation since the source and the target are identical. The figure also shows that correlations are moderate to strong for many pairs, indicating that the trained features may be re-usable, especially for countries that miss ground truth measurements. This result suggests that our model is transferable across countries with similar cultural and geographic backgrounds.

\subsubsection{Backbone network} Our experiments utilized ResNet18 to give a fair comparison with existing research~\cite{han2020lightweight} that reported ResNet18 to give the best result compared to other backbones. To validate the current model choice, we additionally experimented with two different backbone networks -- WideResNet50-2~\cite{zagoruyko2016wide} and DenseNet121~\cite{huang2017densely} -- and report the results below. 
Table~\ref{tab:diff_backbone} demonstrates that there is no significant deviance in performance over the choice of backbone networks, indicating the model is not substantially reliant on a particular backbone network. \looseness=-1

\begin{table}[t!]
\centering
\resizebox{0.99\columnwidth}{!}{
\begin{tabular}{l|ccc}
\toprule
Backbone & ResNet18 & WRN50-2  & DenseNet121\\ \midrule
Tot. Pop.  & 0.8149  &  0.8064  &  0.7788  \\
PP &  0.8212  & 0.7931 & 0.7857 \\  
Ener. Cons. &  0.8553  &  0.8441  & 0.8657\\
GRDP  &  0.4568  & 0.4472 & 0.4864\\\bottomrule
\end{tabular}
}
 \caption{Comparison across different backbone networks.}
 \label{tab:diff_backbone}
\end{table}
\section{Discussion}
For deep learning models to have practical implications for decision-makers, explainability becomes desirable. Here, we provide an interpretation of the computed scores and their implications on estimating economic inequality.

\subsubsection{Model interpretation.} 
\begin{figure}[t!]
\centering
    \begin{subfigure}[b]{0.23\textwidth}
      \includegraphics[width=\textwidth]{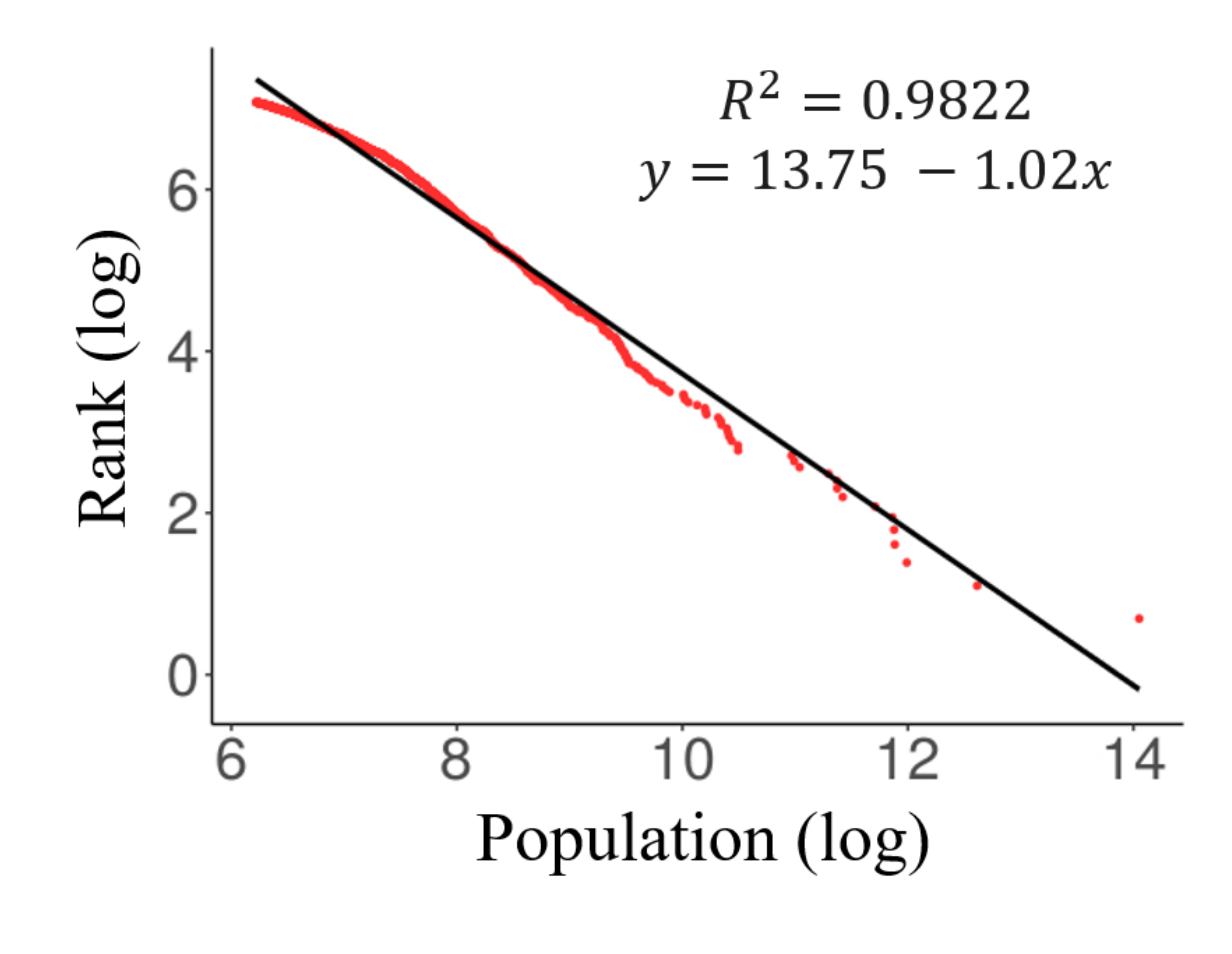}
      \caption{Total population}
    \end{subfigure}
    \begin{subfigure}[b]{0.23\textwidth}
      \includegraphics[width=\textwidth]{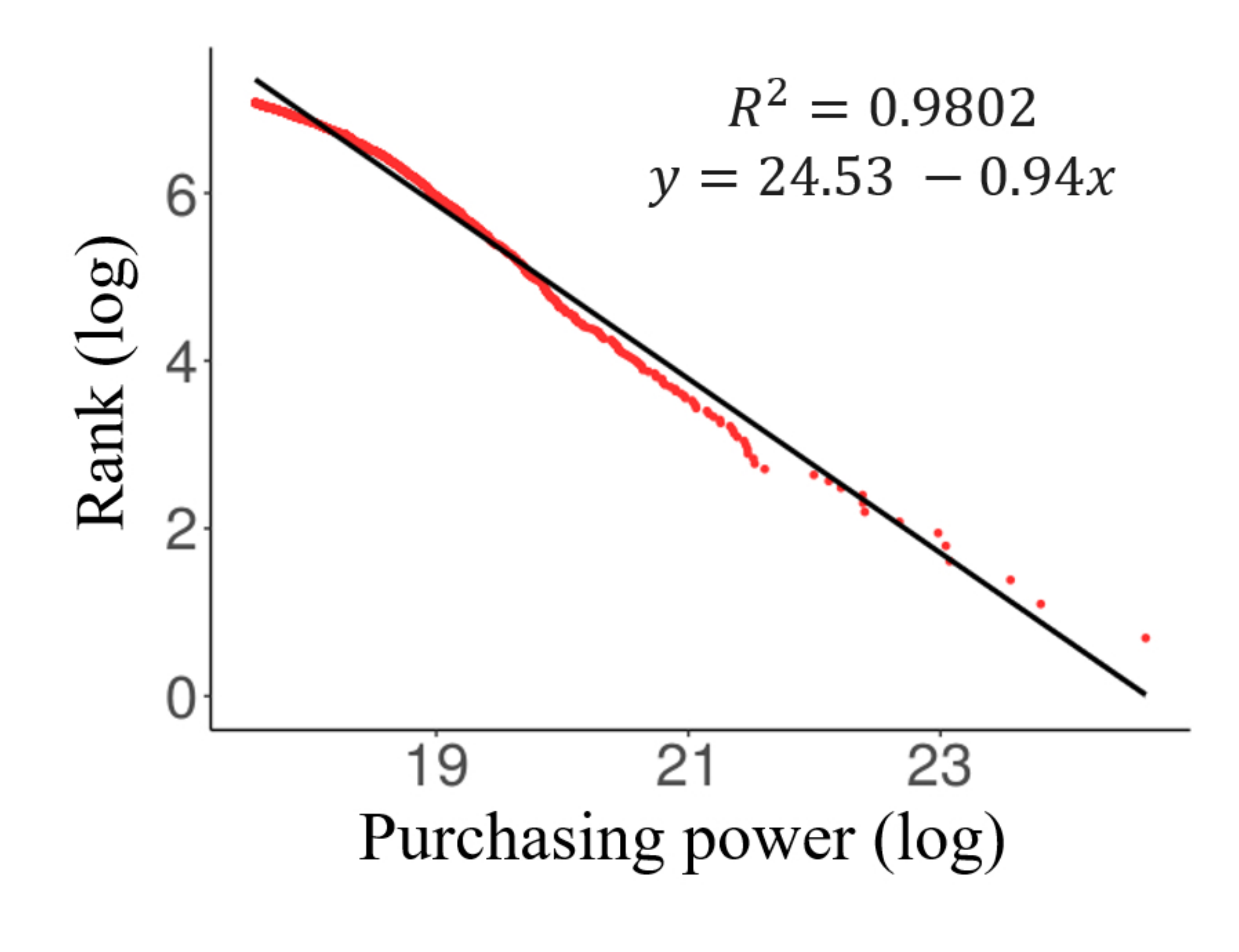}
      \caption{Purchasing power}
    \end{subfigure}   
\caption{Log-rank versus log-size of the district scaling factors based on two economic measures in the Philippines to show the power-law fit. Districts with the top 75\% scaling factors are plotted.}
\label{fig:pareto}
\end{figure}

One of the model's main outputs is $e^{g_{\varphi}(D_i)}$, the multiplicative adjustment that is applied to the hyperlocal scores in Eq~\eqref{eq:scaling}. The distribution of this value follows Zipf's law as shown for the example of the size-rank distribution for two economic indicators in the Philippines (See Figure~\ref{fig:pareto}). These plots show a straight line on a log-log scale, a characteristic fit of the power-law or Zipf's law~\cite{gabaix2009power}.

Zipf's law for cities~\cite{auerbach1913,gabaix1999zipf} is a well-established empirical regularity in economics. The law says that the 2nd largest city is  1/2 the size of the largest city, and the 3rd largest city is 1/3 the size of the largest city. The district scaling factor $g_{\varphi}(D_i)$'s resemblance to the law validates that our model captures this complex economic characteristic well; the district-level economy is not merely a sum of its hyperlocal economies, but their interconnection leads to a disproportionately stronger aggregation effect.

Even more surprising, the power-law holds at the hyperlocal level: the distribution of hyperlocal economic scores $f_\theta(\mathbf{d})$ also shows a straight line in the log-log graph, consistent with the recent study that confirmed the power-law trend at different spatial scales~\cite{mori2020common}.

\begin{figure}[t!]
\centering
\includegraphics[width=0.45\textwidth]{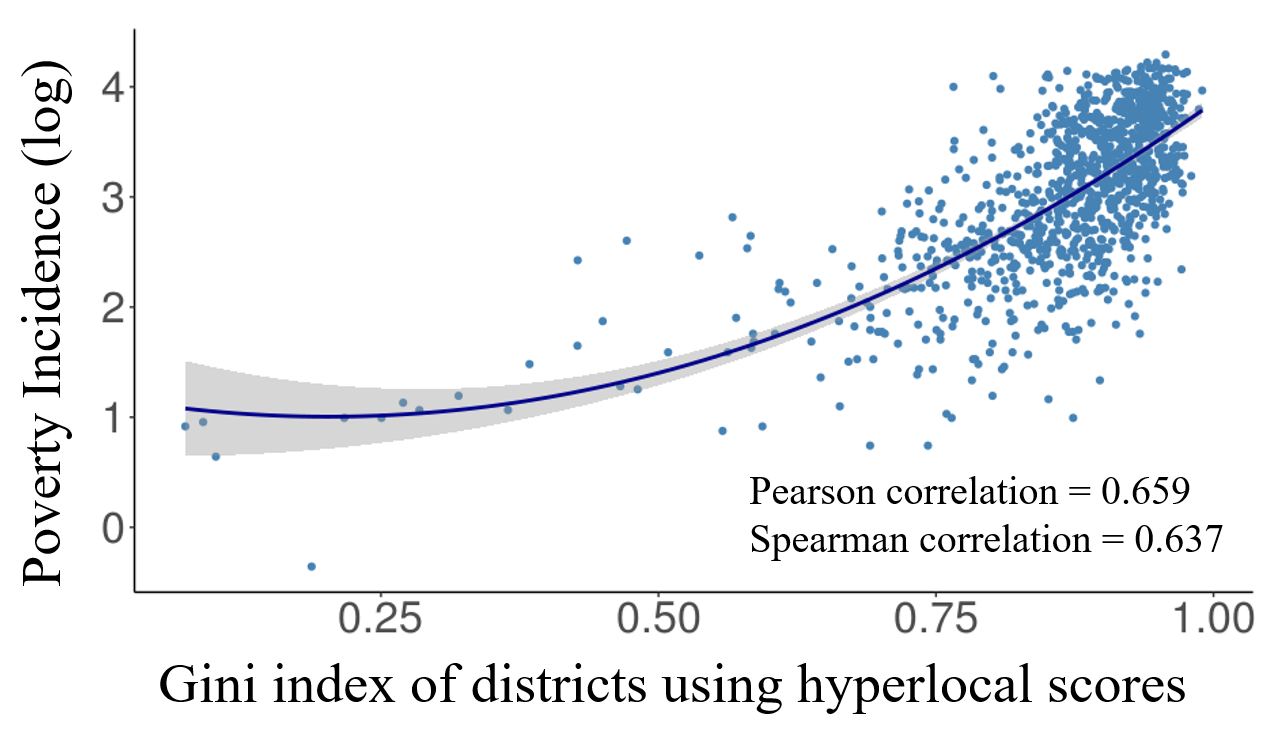} 
\caption{Within-district inequality prediction. The Philippines' poverty incidence is plotted against the Gini index of hyperlocal scores per district, along with the quadratic fit.}
\label{fig:inequality-all}
\end{figure}

\subsubsection{Inequality estimation.} Another application is on economic inequality, a critical challenge for social science research and development policies. Utilizing the model outputs, we can measure inequality at multiple levels.

Inequality is commonly measured by computing the Gini index of economic indicators. Figure~\ref{fig:inequality-all} shows the Gini index of districts $i$ based on the hyperlocal scores $\{f_\theta(\mathbf{d})\  \lvert \ \mathbf{d} \in D_{i} \}$ computed from satellite images of the Philippines. The index is compared against Poverty Incidence, the country's ground-truth poverty indicator by the district. The predicted inequality scores have a relatively high correlation with real data (Pearson 0.659), although no such information was used at training time. This remarkable correlation implies that our within-district estimates can be used to study poverty and inequality at the hyperlocal level, which has not been readily available conventionally. Economic measurements at a granular level are of great potential use for locally tailored policies, as the role of close neighborhoods has recently been recognized as crucial for fighting poverty and addressing inequality problems~\cite{chetty}.

Our multi-level approach for precise hyperlocal measurements can also improve the prediction of national inequality. Table~\ref{table:inequality2} shows that the Gini index calculated from our model shows a high correlation with that published by the World Bank.
Either the hyperlocal information alone (the first row in Table~\ref{table:inequality2}) or aggregated district information (the second row) is not enough to predict the national inequality. The strength of the proposed model is its ability to compute economic indicators at multiple levels, where Gini indices based on the adjusted hyperlocal scores provide the most remarkable correlation with the World Bank data.

\begin{table}[t!]
\centering
\resizebox{0.99\columnwidth}{!}{
\begin{tabular}{lcccc|c}
\toprule
Country &  VT & TH & MY & PH & $\rho_{p}$ \\ \midrule
$f_\theta(\img)$  & 0.468 & 0.360 & 0.571 & 0.514 & 0.656 \\
$e^{g_\varphi(\district)}$  & 0.397 & 0.348 & 0.385 & 0.533 & 0.815 \\
$f_\theta(\img) \cdot e^{g_\varphi(\district)}$ &  0.648 & 0.632 & 0.745 & 0.753 & \textbf{0.942} \\ \midrule
World Bank & 0.357 & 0.364 & 0.410 & 0.444 & 1.000 \\ \bottomrule
\multicolumn{6}{c}{$f_\theta(\img)$ : Hyperlocal score, $e^{g_\varphi(\district)}$ : District scaling factor} \\
\end{tabular}
}
\caption{Prediction of national inequality in purchasing power for four Asian countries: Vietnam (VT), Thailand (TH), Malaysia (MY), and the Philippines (PH). The Gini indices computed by the proposed multi-level model show a high correlation with the World Bank data. ($\rho_{p}$: Pearson correlation)}
\label{table:inequality2}
\end{table}

\section{Conclusion}
This work proposed a new model that utilizes high-resolution geographical images to measure economic indicators over multiple levels (i.e., hyperlocal and district). We first measured the hyperlocal economy captured in a single satellite grid image and then adjusted the cross-district differences via the target district's aggregated features. The model successfully learned the degree of hyperlocal economic development by employing the ordinal regression objective, providing relative orders among satellite grids. The ensemble approach also enabled the model to produce a high-quality representation at the zoom-in view. Our district feature extraction and the district augmentation technique make the model robust against data shortage, a quality that is preferable in many underdeveloped and developing countries. Our multi-level approach shows far advanced prediction performance compared to other baselines. 
 
We showed model's potential in estimating sub-national inequality measures, which is of great interest to urban planners and policymakers. Satellite image-based measurements are robust against data scarcity, and they are light in computation; this means the model can be run repeatedly to generate frequent economic statistics. This ability will benefit countries that require consistent monitoring due to various risks and sustainable growth projections but lack the resources for conventional economic measurements.

\section*{Acknowledgments}
Authors affiliated with KAIST belong to School of Computing. We sincerely thank Eunji Lee and Danu Kim for their valuable feedback on this work. This work was supported in part by the Institute for Basic Science (IBS-R029-C2) and Samsung Electronics. Hyunjoo Yang received the National Research Foundation of Korea grant funded by the South Korean government (No. 2021069641).

\bibliography{aaai22_main}
\end{document}